

\input harvmac

\overfullrule=0pt


\def\no{\noindent}

\def\bs{\bigskip}

\def\bl{\bigl}
\def\br{\bigr}
\def\IR{\relax{\rm I\kern-.18em R}}



\no
hep-th/9207118 \hfill USC-92/HEP-B4

\hfill                    July 1992

\bs\bs

 \centerline {  {\bf HETEROTIC STRING MODELS IN CURVED SPACETIME}
                   {\footnote{$^*$}
  {Research supported in part by DOE, under Grant No. DE-FG03-84ER-40168} } }

\vskip 1.00 true cm

\centerline {ITZHAK BARS }

\bigskip

\centerline {Physics Department}
\centerline {University of Southern California}
\centerline {Los Angeles, CA 90089-0484, USA}


\vskip 1.50 true cm

\centerline{ABSTRACT}
\medskip

We explore the possibility of string theories in only four spacetime
dimensions without any additional compactified dimensions. We show that,
provided the theory is defined in curved spacetime that has a cosmological
interpration, it is possible to construct consistent heterotic string theories
based on a few non-compact current algebra cosets. We classify these
models. The gauge groups that emerge fall within a remarkably narrow range and
include the desirable low energy flavor symmetry of $SU(3)\times SU(2)\times
U(1)$. The quark and lepton states, which come in color triplets and $SU(2)$
doublets, are expected to emerge in several families.

\vfill
\eject


\newsec{Introduction}

During the past year there has been considerable interest in strings
propagating in curved spacetime backgrounds. This was spurred by the fact that
some such models can be formulated as conformally exact current algebra coset
models, or equivalently as gauged WZW models, based on non-compact groups
 \ref\BN{I. Bars and D. Nemeschansky, Nucl. Phys. {\bf 348} (1991) 89.},
and that their geometry describes gravitational singularities of both black
hole
 \ref\WIT{E. Witten, Phys. Rev. {\bf D44} (1991) 314.}
and cosmological types. Before we begin the main technical part of this paper
it is appropriate to make some remarks on why it is interesting to further
study such models. Of course, they provide a setting for  investigating the
very interesting problem of gravitational singularities, but is there more?

{}From experience with string theory we have learned that a conformal field
theory that is a candidate for a ``classical" string vacuum may be expected to
describe the physics at Planck scales. It has been popular to make the
assumption that the string vacuum is flat in four Minkowski dimensions and
that there are additional compactified ``internal" dimensions. In the past all
low energy model building efforts have been based on this unjustified
assumption. Although initially it appeared very promising, the later discovery
of hundreds of thousands of ``vacuua" have diminished the confidence of model
builders. It must be noticed that the multitude of string vacuua occur in the
extra dimensions.

Of course, it was not necessary to assume that the first four dimensions are
flat. Instead, one could imagine a cosmological scenario in which the four
dimensions evolve toward flat spacetime as a function of time. Furthermore,
it was not necessary to assume that there are more than four dimensions.
Recall that extra dimensions appeared historically because the
mathematical consistency of {\it flat} strings required it. But in
curved spacetime, conformal invariance can perfectly well be satisfied in any
dimension, as the non-compact coset models have demonstrated (even just two
dimensions is mathematically consistent). Therefore, it is conceivable that
there are no extra dimensions at all.

String theory is needed to describe physics at very early times or very
short distances near the Planck length. Let us consider a sigma model
formalism which provides a glimpse of the geometry at short distances. Which
features of this geometry can be extrapolated to larger distances? Since there
are a few phase transitions that must be taken into account it is important to
distinguish the features that are likely to be different at large distances
after the phase transitions. First, there is the dilaton which starts out
massless. Since it should not spoil the long range gravitational forces, it
must get a mass near the Planck scale through a phase transition. So far
very little effort has been put into this issue, and it remains as one of the
challenges for string theory. Perhaps this requires understanding
non-perturbative effects. Unfortunately, the present state of affairs allows
us to hide many problems behind this unresolved point. Next, from experience
with grand unified theories one also knows that phase transitions associated
with gauge forces, through the mechanism of inflation, can explain why the
universe is homogeneous and isotropic.

So, the universe (as described by the sigma model metric) need not start out
homogeneous and isotropic or flat in four dimensions. It would be sufficient
to start out with a {\it part of the universe} which is expanding in four
dimensions, and that by the time its size reaches a few Planck scales it
approaches a flat universe. If this part of the universe undergoes inflation
it may describe our observed universe. The background geometries provided by
the non-compact coset models include such geometries in 3d and 4d (see e.g.
 \ref\BASglo{I. Bars and K. Sfetsos, ``Global Analysis of New Gravitational
Singularities in String and Particle Theories'',
USC-92/HEP-B1 (hep-th/9205037), to appear in Phys. Rev. D (1992).}
 \ref\BASexa{I. Bars and K. Sfetsos, ``Conformally Exact Metric and Dilaton
in String Theory on Curved Spacetime", USC-92/HEP-B2
 (hep-th 9206006), to appear in Phys. Rev. D (1992).
         \semi Konstadinos Sfetsos, ``Conformally Exact Results for
$SL(2,\IR)\times SO(1,1)^{d-2}/SO(1,1)$ Coset Models", USC-92/HEP-
S1, (hep-th 9206048) }).
 In addition, {\it heterotic} string models with such backgrounds predict
gauge fields and a spectrum of matter that provides candidates for the low
energy quarks and leptons. We know that the forces associated with gauge
fields and self couplings of matter could explain the mechanism for mass
generation. So, we may defer the mass generation problem to energies well
below the Planck scale.

Therefore we may consider a scenario in which there are only three space and
one time dimensions. Then the conformal string theory must be in curved
spacetime and is designed to satisfy the conditions of exact conformal
invariance. The geometry at the Planck length is not necessarily homogeneous
or isotropic. At least some bundle of geodesics (that represent the early
evolution of part of the universe that gets later inflated) migrate to regions
of flat spacetime within a few units of Planck time, perhaps exponentially (as
in the $d=3,4$ non-compact models \BASglo\BASexa\ ). The gauge and matter
fields of the heterotic theory can survive to low energies through the
mechanism of gauge symmetries and chiral symmetries. Some of this ``low energy
matter" will become all of the matter in the inflated universe. Therefore,
such a heterotic string theory can be used to at least classify the particles
in multiplets of the symmetry group and compare to the known low energy
classification of quarks and leptons.

The initiation of such a program is one of the purposes of the present paper.
We will classify the heterotic string models in just four dimensions that can
be constructed as exact conformal theories based on non-compact groups. We
find that the list of such models is rather short. We will be able to extract
the gauge symmetry content of these models and show that the possible gauge
groups fall within a remarkably narrow range, and always include the desirable
low energy symmetries. This approach does not explain why we live in four
dimensions, and of course the program can be carried out also with additional
compact dimensions. But it seems very interesting to find out what kinds of
results emerge if there are in fact only four dimensions.

There exists by now a few models of strings propagating in curved spacetime
that are in principle solvable due to the fact that they are formulated as
non-compact current algebra cosets based on non-compact groups. The
classification of the cosets $G/H$ that yield a single time coordinate \BN\ is
known
 \ref\IBclass{I. Bars, ``Curved Space--Time Strings and Black Holes'',
in Proc. of {\it XX$^{th}$ Int. Conf. on Diff.
Geometrical Methods in Physics}, Eds. S. Catto and A. Rocha, Vol.2, p.695,
(World Scientific, 1992).\semi
P. Ginsparg and F. Quevedo, ``Strings on Curved Space-Times:
Black Holes, Torsion, and Duality, LA-UR-92-640.
            \semi See also \BASglo\ }.
The cosets that lead to models in four curved spacetime dimensions
($D=4$) always include $SO(d-1,2)/SO(d-1,1)$ for $d\le 4$. In this paper we
will assume that there are no more than $D=4$ dimensions and therefore use
only $SO(d-1,2)$ for $d=2,3,4$. For $D=d=4$ there are no other bosonic
coordinates.  When $d\le 3$, then $D-d=4-d$ additional bosonic coordinates are
supplied by taking direct products with other groups (including space-like
$U(1)$ or $\IR$ factors) and then gauging an appropriate subgroup.
Furthermore, we include in our list the possibility of a time-like bosonic
coordinate and denote it by a factor of $T$ instead of $\IR$. All
possibilities
 are listed in Table-1 in the column labelled ``right movers".


$$\vbox   {    \tabskip=0pt \offinterlineskip
\halign to 388pt  { \vrule# & \strut # &
 #\hfil &\vrule# \ \ &  #\hfil &\vrule# \ \  &  #\hfil &\vrule#
\tabskip =0pt
\cr \noalign {\hrule}
&& \# && left movers with N=1 SUSY && right movers &
\cr\noalign{\hrule}
\noalign {\smallskip}
&&$ 1 $&&${SO(3,2)_{-k}\times SO(3,1)_{1}/ SO(3,1)_{-k+1}} $&
        &$  {SO(3,2)_{-k}/ SO(3,1)_{-k} }  $&
\cr\noalign{\hrule}
&&$ 2 $&&$ {SL(2,\IR)_{-k_1}\times SL(2,\IR)_{-k_2}\times SO(3,1)_1\over
        SL(2,\IR)_{-k_1-k_2+2}}\times \IR $&
        &$  {SL(2,\IR)_{-k_1}\times SL(2,\IR)_{-k_2}\over SL(2,\IR)_{-k_1-k_2}}
 \times \IR   $&
\cr\noalign{\hrule}
&&$ 3 $&&$ \bl (SO(2,2)_{-k}\times SO(3,1)_1/SO(2,1)_{-k+2}\br)\times \IR $&
       &$ \bl (SO(2,2)_{-k}/SO(2,1)_{-k}\br)\times \IR $&
\cr\noalign{\hrule}
&&$ 4 $&&$ SL(2,\IR)_{-k}\times SO(3,1)_1\times \IR $&
        &$ SL(2,\IR)_{-k}\times \IR $&
\cr\noalign{\hrule}
&&$ 5 $&&$ {SL(2,\IR)_{-k_1}\times SL(2,\IR)_{-k_2}\times SO(3,1)_1 \over
          T\times \IR } $&
        &$ {SL(2,\IR)_{-k_1}\times SL(2,\IR)_{-k_2} / (T\times \IR) }   $&
\cr\noalign{\hrule}
&&$ 6 $&&$ {SL(2,\IR)_{-k_1}\times SU(2)_{k_2}\times SO(3,1)_1 / \IR^2} $&
        &$  {SL(2,\IR)_{-k_1}\times SU(2)_{k_2}/ \IR^2}  $&
\cr\noalign{\hrule}
&&$ 7 $&&$ {(SL(2,\IR)_{-k}\times \IR^2\times SO(3,1)_1)/ \IR }  $&
        &$ {(SL(2,\IR)_{-k}\times \IR^2 )/ \IR }  $&
\cr\noalign{\hrule}
&&$ 8 $&&$ T\times \IR^3\times SO(3,1)_1  $&&$ T\times \IR^3   $&
\cr\noalign{\hrule}
&&$ 9 $&&$ T\times {SU(2)_{k_1}\times SU(2)_{k_2}\times SO(3,1)_1
          \over SU(2)_{k_1+k_2+2} }  $&
        &$ T\times {SU(2)_{k_1}\times SU(2)_{k_2}\over SU(2)_{k_1+k_2} }  $&
\cr\noalign{\hrule}
&&$ 10 $&&$ T\times SO(4)_{k}\times SO(3,1)_1/ SU(2)_{k+2}   $&
        &$ T\times SO(4)_{k}/ SU(2)_{k}   $&
\cr\noalign{\hrule}
&&$ 11  $&&$ T\times SU(2)_{k}\times SO(3,1)_1 $&
        &$ T\times SU(2)_{k} $&
\cr\noalign{\hrule}
&&$ 12 $&&$ {(T\times\IR\times SU(2)_k\times SO(3,1)_1)/ \IR}  $&
        &$ {(T\times\IR\times SU(2)_{k})/ \IR} $&
\cr\noalign{\hrule}
&&$ 13 $&&$ {(T\times\IR\times SL(2,\IR)_{-k}\times SO(3,1)_1)/ T}  $&
        &$ {(T\times\IR\times SL(2,R)_{-k})/ T} $&
\cr\noalign{\hrule}
&& \multispan5 Table-1. Current algebraic description of left movers and
right movers.\hfill &
\cr\noalign{\hrule}
}}$$
For brevity we used $\IR$ where we could have used either $\IR$ or $U(1)$.
Case 3 is obtained from case 2 in the limit $k_1=k_2=k$, while case 4 is the
$k_1=k,\ k_2=\infty$ limit of either case 2 or 5. Similarly, cases 10,11 are
limits of case 9. Furthermore, case 8 may be considered the large $k$ limit of
case 11. This last case has unique properties in that its geometry is
flat, homogeneous and isotropic (modulo boundary conditions on the $\IR^3$
factor). One may also notice that cases 9-13 are analytic continuations of
cases 3-7. We have listed all these limits or analytic continuations
separately because they lead to different gauge groups as will be seen in
Table-2 below.

For the numerator factor $T$ in cases 8-13 we allow a background charge $Q_0$.
The background charge  for the time-like coordinate contributes
$c_T=(1+12Q_0^2)$ to the Virasoro central charge, and this quantitity is
always larger than one. Similarly, every space-like coordinate associated with
the factors of $\IR$ in the numerators may be allowed to have a non-trivial
background charge $Q$. This contributes $c_{\IR}=(1-12Q^2)$ for a space-like
coordinate, and is always less than one and positive. In the following, to
keep our expressions simple, we will assume that $Q=0$. A non-zero $Q$ makes
no difference for the discussion below, but we will indicate separately the
changes that occur at intermediate steps.

The cases 5,6,7,12,13 which contain a $T$ or $\IR$ factor in the
denominator may further be generalized by multiplying both numerator and
denominator by a factor $\IR^n$. What this implies is that there are many
possible ways of gauging the $\IR$ and/or $T$ factors by taking linear
combinations. This may lead to models that are different, however this
generalization does not change the results given in Table-2 at all.

The heterotic string will have a supersymmetric left-moving sector and a
non-supersymmetric right-moving sector. The cosets above describe the four
dimensional space-time part of the right-moving sector.  This contributes
$c_R(4D)$ toward the Virasoro central charge. After we analyse the central
charge of the supersymmetric left movers and fix it to be $c_L=15$ in only four
dimensions, we will see that $c_R(4D)$ will be fixed to some value less
than $26$. Therefore, for the mathematical consistency of the theory, we must
require that the right moving sector contains an additional ``internal" part
which makes up for the difference, i.e. $c_R(int)+c_R(4D)=26$. One of the
aims of this paper is to compute $c_R(int)$ in each model and then find gauge
symmetry groups that precisely give this value. This procedure will allow us
to discover the gauge symmetries that are possible in these curved spacetime
string models.

To construct a heterotic string we introduce four left moving coset fermions
$\psi^\mu$ that are classified under $H$ as $G/H$ and form a $N=1$
supermultiplet together with the four bosons. The construction of the action
that pocesses the superconformal symmetry is done along the lines of
 \ref\BAShet{I. Bars and K. Sfetsos, Phys. Lett. {\bf 277B} (1992) 269.}.
The left moving fermions $\psi^\mu$ are coupled to the gauge bosons in $H$. In
the Hamiltonian language, the left moving stress tensor is expressed in the
form of current algebra cosets
 \ref\KS{Y. Kazama and H. Suzuki, Nucl.Phys. B234 (1989) 232.}
 \ref\IBhe{I. Bars, Nucl. Phys. {\bf B334} (1990) 125. }
as listed in Table-1, where $SO(3,1)_1$ represents the fermions.

This algebraic formulation allows an easy computation of the Virasoro
central charges for left movers $c_L$ as well as the right movers $c_R(4D)$.
For a consistent theory we must set $c_L=15$. This condition puts restrictions
on the various central extensions $k$ and/or background charges $Q_0,Q$, as
listed in Table-2. After inserting these in $c_R(4D)$ we find the deficit from
the critical value of 26, i.e. $c_R(int)=26-c_R(4D)$. As seen in the table,
the resulting values for $c_R(int)$ fall within a narrow range. For case
2 or 3 it is possible to change the central charge within the range $11{1\over
2}<c_R(int) <13$ by varying  $k_1+k_2$. Similarly, the corresponding range for
cases 9,10 is $12{1\over 2}<c_R(int)<13$. For the remaining cases it is not
possible to change $c_R(int)$ by using the remaining freedom with the $k's$.


$$\vbox   {    \tabskip=0pt \offinterlineskip
\halign to 377pt  { \vrule# & \strut # &
 #\hfil &\vrule# \ \ &  #\hfil &\vrule# \ \  &  #\hfil &\vrule#
 \  & #\hfil &\vrule#
\tabskip =0pt
\cr \noalign {\hrule}
&& \# && conditions for $c_L=15 $ && $c_R(int)$ && gauge group, right movers &
\cr\noalign{\hrule}
\noalign {\smallskip}
&&$ 1 $&&$ k=5 $&&$ 11 $&&$ (E_7)_1\times SU(5)_1 $&
\cr\noalign{\hrule}
&&$ 2 $&&$ k_1-2={k_2-2\over 2}(-1+\sqrt {3k_2\over 3k_2-8})   $&
        &$ 13-\delta $&&$ \delta={12\over (k_1+k_2-4)(k_1+k_2-2)}  $&
\cr\noalign{\hrule}
&&$ 3 $&&$ k=3   $&&$ 11{1\over 2} $&
        &$ (E_7)_1\times SU(3)_1\times SU(2)_2\times U(1)_1  $&
\cr\noalign{\hrule}
&&$ 4 $&&$ k=8/3   $&&$ 13  $&&$ (E_8)_1\times SO(10)_1  $&
\cr\noalign{\hrule}
&&$ 5 $&&$ k_1={8k_2-20\over 3k_2-8},\ \ k_1,k_2>{8\over 3}  $&
        &$ 13 $&&$ (E_8)_1\times SO(10)_1 $&
\cr\noalign{\hrule}
&&$ 6 $&&$k_1={8k_2+20\over 3k_2+8}, \ k_2=1,2,3,\cdots $&
        &$ 13  $&&$ (E_8)_1\times SO(10)_1 $&
\cr\noalign{\hrule}
&&$ 7 $&&$ k={8/3} $&&$ 13  $&&$ (E_8)_1\times SO(10)_1  $&
\cr\noalign{\hrule}
&&$ 8 $&&$ Q_0^2={3\over 4} $&&$ 13  $&&$ (E_8)_1\times SO(10)_1  $&
\cr\noalign{\hrule}
&&$ 9 $&&$ Q_0^2={3\over 4}+{1\over 2}({1\over k_1+2}+
                 {1\over k_2+2}-{1\over k_1+k_2+4})   $&
        &$ 13-\delta $&&$ \delta={12\over (k_1+k_2+4)(k_1+k_2+2)}  $&
\cr\noalign{\hrule}
&&$ 10 $&&$ Q_0^2={3(k+3)\over 4(k+2)},\ (e.g.\ \ k=1)   $&&$ 12{1\over 2}  $&
        &$ (E_8)_1\times SU(3)_1\times SU(2)_2\times U(1)_1  $&
\cr\noalign{\hrule}
&&$ 11 $&&$ Q_0^2={3k+8\over 4(k+2)},\ k=1,2,3,\cdots $&
        &$ 13 $&&$ (E_8)_1\times SO(10)_1  $&
\cr\noalign{\hrule}
&&$ 12 $&&$ Q_0^2={3k+8\over 4(k+2)},\ k=1,2,3,\cdots $&&$ 13  $&
        &$ (E_8)_1\times SO(10)_1  $&
\cr\noalign{\hrule}
&&$ 13 $&&$ Q_0^2={3k-8\over 4(k-2)} $&&$ 13 $&&$ (E_8)_1\times SO(10)_1 $&
\cr\noalign{\hrule}
&& \multispan7 Table-2. Conditions for $c_L=15$ and examples of symmetries
that give $c_R=26$. &
\cr\noalign{\hrule}
}}$$

At this point we mention the effect of a non-zero background charge $Q$
for the space-like factor $\IR$ in cases 2,3,4 and 7. The formula for
$c_R(int)$ that is listed in the table for case 2 remains the same, but the
conditions on the $k's$ change slightly. The new conditions take the same form
as cases 9,10,11 and 12 as listed in the table respectively, except for the
analytic continuations $Q_0^2\rightarrow -Q^2$ and $k_i\rightarrow -k_i$.
However, since $0<Q^2<{1\over 12}$, the values of the new $k's$ in cases 2,5
must remain within a narrow range of those already fixed in Table-2. Only for
cases 2,3,4 this has an effect on $c_R(int)$. For example, for case 3 we get
$2.9\le k\le 3$ (instead of $k=3$) and  $11.24 < c_R(int)<11{1\over 2}$
(instead of $11{1\over 2}$). The presence of a non-zero $Q$ does not change the
discussion that follows.

The value of $c_R(int)=13$ that occurs for most of the cases is the same as
the deficit for the popular heterotic string models that have four flat
dimensions plus compactified dimensions described by a $c=9$, $N=2$
superconformal theory (i.e. $4+9+13=26$). Hence, for these cases, the
appearance of $(E_8)_1\times SO(10)_1$ as the gauge group has precisely the
same explanation as the usual approach. For the remaining cases we give an
example of a gauge symmetry that will make up the deficit, as listed in Table-
2. Other gauge groups are clearly possible just on the basis of $c_R(int)$.
For example, for case 1 one can have $SO(22)_1,\ \ \ (E_8)_1\times SU(4)_1, \ \
\ (E_7)_1\times SU(5)_1 ,\ \ \ (E_7)_1\times SU(3)_1\times SU(2)_1\times U(1)_1
, \ \ \ (E_6)_1\times SO(10)_1 $, etc., as given in \BAShet .

The gauge symmetry is associated with a conformal theory of right movers. This
additional part of the action may be constructed from right moving free
fermions with appropriate boundary conditions, or by using other devices that
are quite familiar. We can think of this part as another current algebra
associated with the gauge group, and with the central extensions that are
given in Table-2. This final step completes the action for the model. For the
complete action for case 1, see \BAShet . Further discussion of the model is
required to determine the symmetries consistent with modular invariance. At
this stage it is encouraging to note that {\it the desirable low energy
symmetries, including $SU(3)\times SU(2)\times U(1)$, are contained in these
curved space string models that have only four dimensions}.

The special property of the models constructed in this paper is that they can
be further investigated by using current algebra techniques. The simplest
model is case 8, since it is essentially flat, its quantum theory reduces
to the manipulation of harmonic oscillators. For the remaining models the
spectrum of low energy particles is obtained by computing the quadratic
Casimir operators of the non-compact groups that define the model. The
computation of the spectrum will be reported in a future publication. Since
the flavor groups such as $SU(3)\times SU(2)\times U(1)$ or $SU(5),\ SO(10)$,
etc. appear at level 1, it is already evident that the quark and lepton type
of matter will appear in triplets and doublets respectively.

An interesting question is how the repetition of the families will come about?
In the traditional approach that includes compactified dimensions, the number
of families is related to indices, such as Betti numbers, of the compactified
space. In the present case, the four dimensional geometry has many new and
interesting properties, such as duality, different topological sectors, etc. as
seen in the global analysis of \BASglo . Therefore, we may expect that the
repetition of families may have something to do with these properties. The
repetition will show up in the algebraic approach by the number of distinct
ways that it is possible to satisfy the on-mass-shell conditions for the same
quark or lepton quantum number (e.g. different representations of the
non-compact group). When this criterion is applied to the flat case 8, we see
that the presence of the background charge $Q_0$ allows two distict states to
be associated with the same conformal dimension, thus leading to two families.
Therefore, it is quite possible that the repetition of families may arise from
the four dimensional geometry alone as described above.

We want to point out another possible source of family replication. It
may be feasible to interpret part of the gauge group as a ``family group", as
it was done in the days before string theory. If this latter alternative is
utilised for family replication, then one might quit the idea of ``hidden
sectors" attributable to groups such as $E_8$, and instead adopt a version of
the gauge group which has complex representations. For example, in case 1, one
of the possibilities that give $c_R(int)=11$ was $(E_6)_1\times SO(10)_1$ ,
which has complex representations.

During the past year there has been many investigations [2-17]
\nref\BSthree{I. Bars and K. Sfetsos, Mod. Phys. Lett. {\bf A7} (1992) 1091.}
\nref\CRE{M. Crescimanno, Mod. Phys. Lett. {\bf A7} (1992) 489.}
\nref\HOHO{J. B. Horne and G. T. Horowitz, Nucl. Phys. {\bf B368} (1992) 444.}
\nref\FRA{E. S. Fradkin and V. Ya. Linetsky, Phys. Lett. {\bf 277B} (1992) 73.}
\nref\ISH{N. Ishibashi, M. Li, and A. R. Steif,
         Phys. Rev. Lett. {\bf 67} (1991) 3336.}
\nref\HOR{P. Horava, Phys. Lett. {\bf 278B} (1992) 101.}
\nref\RAI{E. Raiten, ``Perturbations of a Stringy Black Hole'',
         Fermilab-Pub 91-338-T.}
\nref\GER{D. Gershon, ``Exact Solutions of Four-Dimensional Black Holes in
         String Theory'', TAUP-1937-91.}
that explored
the geometry of the sigma model-like action associated with some of these
models. While the geometry for $d=2$ is interpreted as a black hole
the singularity structure for $d=3,4$ is considerably more involved and
interesting. For example, for $d=3$ a global analysis of the manifold shows
that there are two topologically distinct sectors that can be pictured as the
``pinched double trousers" or the ``double saddle" \BASglo .
Furthermore, the time dependent backgrounds that emerge allow for cosmological
interpretations.
These results were initially obtained at the semi-classical level using the
lagrangian method (in patches of the geometry). More recently, fully quantum
mechanical results were obtained by using
conformally exact current algebra methods in a Hamiltonian formalism \BASexa .
The algebraic method simultaneously yields the full global geometry, as was
illustrated for $d=2,3,4$.
 Furthermore, the heterotic and type-II supersymmetric versions of
these models were investigated and their conformally exact global geometry
determined. By now the global geometry of all the above models have essentially
been completed in \BASglo\BASexa
 \ref\BASslsu{I. Bars and K. Sfetsos, $SL(2,\IR)\times SU(2)/\IR^2$ String
Model in Curved Spacetime and Exact Conformal Results, USC-92/HEP-B3 (hep-th
920800x) }. The cases not covered directly in these references can be
obtained with analytic continuation techniques.

Are there additional models beyond the ones listed in the tables? Undoubtedly
there are more, but they may not have the virtue of being solvable like the
present ones that have a current algebra formulation. It is, of course,
possible to imagine perturbations of the present models that may be formulated
in the current algebra language and yield solvable cases. Such perturbations
will tend to change the formulas for the central charges and it would be
interesting to investigate how stable is $c_R(int)$ against these
perturbations and how the gauge group is affected by them. On the basis of
Table-2 it seems that one cannot wander too far away from $c_R(int)=13$.
Therefore the desirable low energy flavor symmetries are likely to remain.
These are interesting questions that should be investigated.

We have argued that it is interesting to consider the possibility of heterotic
string theories in only four spacetime dimensions and no additional
compactified dimensions. This is possible only in curved spacetime, and
such a string can be imagined to describe the very small distances or very
early times in the Universe.

\listrefs
\end